\documentclass[twocolumn,showpacs,aps,prl,superscriptaddress,amssymb,floatfix,preprintnumbers]{revtex4}
\usepackage{graphicx,rotating}
\usepackage{dcolumn}
\usepackage{epsfig}
\usepackage{multirow}

\input babarsym

\long\def\inst#1{\par\nobreak\kern 4pt\nobreak
    {\it #1}\par\vskip 10pt plus 3pt minus 3pt}

\begin{document}

\preprint{  \babar-PUB-06/055}   \preprint{SLAC-PUB-12075}

\title{
\Large \bf \boldmath
Observation of an Excited Charm Baryon \ocs\ Decaying to $\ocz\g$}

%
\author{B.~Aubert}
\author{M.~Bona}
\author{D.~Boutigny}
\author{F.~Couderc}
\author{Y.~Karyotakis}
\author{J.~P.~Lees}
\author{V.~Poireau}
\author{V.~Tisserand}
\author{A.~Zghiche}
\affiliation{Laboratoire de Physique des Particules, IN2P3/CNRS et Universit\'e de Savoie,
 F-74941 Annecy-Le-Vieux, France }
\author{E.~Grauges}
\affiliation{Universitat de Barcelona, Facultat de Fisica, Departament ECM, E-08028 Barcelona, Spain }
% ...
\collaboration{The \babar\ Collaboration}
\noaffiliation

\date{\today}

\begin{abstract}

\indent We report the first observation of an excited singly-charmed baryon $\Omega_{c}^{*}$ ($css$) in the radiative decay $\Omega_{c}^{0} \gamma$, where the $\Omega_{c}^{0}$ baryon is reconstructed in the decays to the final states $\Omega^{-}\pi^{+}$, $\Omega^{-}\pi^{+}\pi^{0}$, $\Omega^{-}\pi^{+}\pi^{-}\pi^{+}$, and $\Xi^{-}K^{-}\pi^{+}\pi^{+}$. This analysis is performed using a dataset of 230.7 fb$^{-1}$ collected by the BABAR detector at the PEP-II asymmetric-energy $B$ Factory at the Stanford Linear Accelerator Center. The mass difference between the $\Omega_{c}^{*}$ and the $\Omega_{c}^{0}$ baryons is measured to be 70.8~$\pm$~1.0~(stat)~$\pm$~1.1~(syst) MeV/c$^{2}$. We also measure the ratio of inclusive production cross sections of $\Omega_{c}^{*}$ and $\Omega_{c}^{0}$ in e$^{+}$e$^{-}$ annihilation. 

\end{abstract}
\maketitle

\setcounter{footnote}{0}

The production of charm baryons is largely unexplored and provides an interesting environment to study the dynamics of quark-gluon interactions. All singly-charmed baryons  having zero orbital angular momentum have been discovered~\cite{PDG}, except for the $J^P={3\over2}^+$ $css$ state, denoted as \ocs. A non-relativistic QCD effective field theory calculation predicts the difference between the mass of \ocs($M_{\ocs}$) and the mass of \ocz($M_{\ocz}$), $\Delta M$, to be between 50 and 73 \mevcc~\cite{Mathur}. A lattice QCD calculation gives $\Delta M = 94 \pm 10$ \mevcc~\cite{1}. New quadratic baryon mass relations predict a mass of $M_{\ocs}$ = $2767\pm7$\mevcc~\cite{2}, and several other predictions for $M_{\ocs}$ exist around 2770 \mevcc~[5-11], implying $\Delta M = 70 - 75$ \mevcc.

Here we report the observation of an excited baryon \ocs produced inclusively in $\epem \to \ocs X$ processes, where $X$ denotes the rest of the event. We measure the mass difference, $\Delta M$, and the ratio of the production cross section of $\epem \to \ocs X $ relative to $\epem \to \ocz X $. Throughout this paper, for any given mode, the corresponding charge conjugate reaction is also implied.

      The data used in this analysis were collected with the \babar\ detector at the \pep2\ asymmetric-energy \epem\ storage rings. The dataset corresponds to an integrated luminosity of 209.1 \invfb\ collected at a center-of-mass (CM) energy of $\sqrt{s}$ = 10.58 GeV, near the peak of the \FourS\ resonance, and 21.6 \invfb\ collected approximately 40 MeV below the \FourS\ mass.

    The \babar\ detector is described elsewhere~\cite{ref:babar}. Charged tracks are reconstructed with a five-layer, double-sided silicon vertex tracker (SVT) and a 40-layer drift chamber (DCH) with a helium-based gas mixture, placed in a 1.5-T uniform magnetic field produced by a superconducting solenoidal magnet. Kaons, pions and protons are identified using likelihood ratios calculated from the ionization energy loss (\dedx) measurements in the SVT and DCH, and from the observed pattern of Cherenkov light in an internally reflecting ring imaging detector. Photons are identified as isolated electromagnetic showers in a CsI(Tl) electromagnetic calorimeter (EMC). Large samples of Monte Carlo (MC) simulated data are used for determination of signal detection efficiencies and for the optimization of the selection criteria. These are generated using JETSET~\cite{jetset} and the detector response is simulated with GEANT4~\cite{G4}. 

 The \ocs\ candidate is identified through its radiative decay, \ocstg, where the \ocz\ is reconstructed exclusively in the following four decay modes, which are expected to provide the best signal-to-background ratio: 
\begin{center}
\begin{itemize}
\item \octpi, \ommtolzk\         ~~~~~~~~~~~~~  (O1)
\item \octpipi, \ommtolzk\       ~~~~~~~~~~       (O2)
\item \octpipipi, \ommtolzk\     ~~~~~~   (O3)
\item \octckpipi, \casmtolzp\    ~~~~~~    (C1)
\end{itemize}
\end{center}

The labels in parentheses to the right of each decay mode designate the four final states of the \ocz decay. 

A \lztoppi\ candidate is reconstructed by identifying a proton track, combining it with an oppositely-charged track identified as a \pim, and fitting the tracks to a common vertex. Here and throughout this analysis, all reconstructed baryon candidates are required to have an acceptable $\chi^{2}$ from the vertex fit. The flight distance of each \lz\ candidate between its decay vertex and that of its parent (\omm or \casm) is required to be greater than 0.30 cm. The \lztoppi\ signal is fitted using a sum of two Gaussian functions with a common mean. The signal region is defined by $|M_{p\pi^{-}} - M_{\Lambda}| < 3.8 $\mevcc\ ($\approx 2 \sigma_{\rm {RMS}}$), where $M_{\Lambda}$ is the fitted peak position of the \lz\ and $\sigma_{\rm {RMS}}$ is defined by $ \sigma_{\rm {RMS}}^{2} \equiv f_{1} \sigma_{1}^{2} + f_{2} \sigma_{2}^{2},$ where $f_{1}$ and $f_{2}$ are the fractions of the two Gaussian functions, and $\sigma_{1}$ and $\sigma_{2}$ are the two corresponding widths as obtained from the fit. The reconstructed \lz\ candidate is then combined with an identified $K^{-}$ (\pim ) to form an \omm (\casm) candidate. The \lz\ and the $K^{-}$ (\pim ) tracks are fitted to a common vertex, and the flight distance of each \omm\ or \casm\ candidate between its decay vertex and that of its parent (\ocz) is required to be greater than 0.25 cm. Mass windows of $| M_{\Lambda K^{-}} - M_{\Omega^{-}}| <$ 5.2 \mevcc ($\approx$ 2$\sigma_{\rm {RMS}}$) and $| M_{\Lambda \pi^{-}} - M_{\Xi^{-}}| <$ 6.0 \mevcc~($\approx$~2$\sigma_{\rm {RMS}}$)~ are used to select \ommtolzk and \casmtolzp candidates, respectively, where $M_{\Omega^{-}}$ and $M_{\Xi^{-}}$ represent the fitted peak positions of \omm\ and \casm.

  For the decay mode O2, the \piz\ candidates are reconstructed by combining two photons. To enhance the \piz\ signal over combinatorial background, we require photons to have a minimum energy of 80 MeV in the laboratory frame, to have a lateral shower shape consistent with that of a photon and to be well-separated from other tracks and clusters in the EMC. We require $| M_{\g\g} - M_{\piz} | < 12.5$ \mevcc (2.5$\sigma$), where $M_{\piz}$ is the fitted peak position of the invariant mass of the two photons. 

  For decays O1, O2, O3, the reconstructed \omm\ is combined with a (\pip, \pippiz, \pippimpip) to form an \ocz, and fitted to a common vertex. For C1, the reconstructed \casm\ is combined with an identified \Km\ and two \pip\ tracks and fitted to a common vertex. The invariant mass of reconstructed \ocz\ candidates is required to lie within $\pm2.5~\sigma_{\rm {RMS}}$ of the central fitted value. The mass resolution is $\sigma_{\rm {RMS}}\approx$ 6 \mevcc\ for O1, O3, and C1, and  $\sigma_{\rm {RMS}} \approx$ 13 \mevcc\ for O2. The resolution in O2 is dominated by the measurement of the photon energies from the \piz decay.

  An \ocs\ candidate is formed by combining a reconstructed \ocz\ with a photon, applying the same photon selection requirements listed above for photons from \piz\ decay. For O2, it is required that the photon is not one of the \piz\ daughters.

Though eliminating most \ocs\ baryons from $B$ decays, the requirement that the scaled momentum of \ocs\ candidates, ($x_{p}({\ocs})$), be greater than 0.5 significantly reduces combinatorial background from $\epem \to q\bar{q}$ (where \q = \u, \d, \s). The scaled momentum is defined as $x_p=p^*/p^*_{max}$, where $p^*$ is the reconstructed momentum in the CM frame and $p^*_{max}=\sqrt{s/4-M^2}$, with $M$ being the mass of the particle.

\begin{figure}
\begin{center}
\includegraphics[height=5.8in]{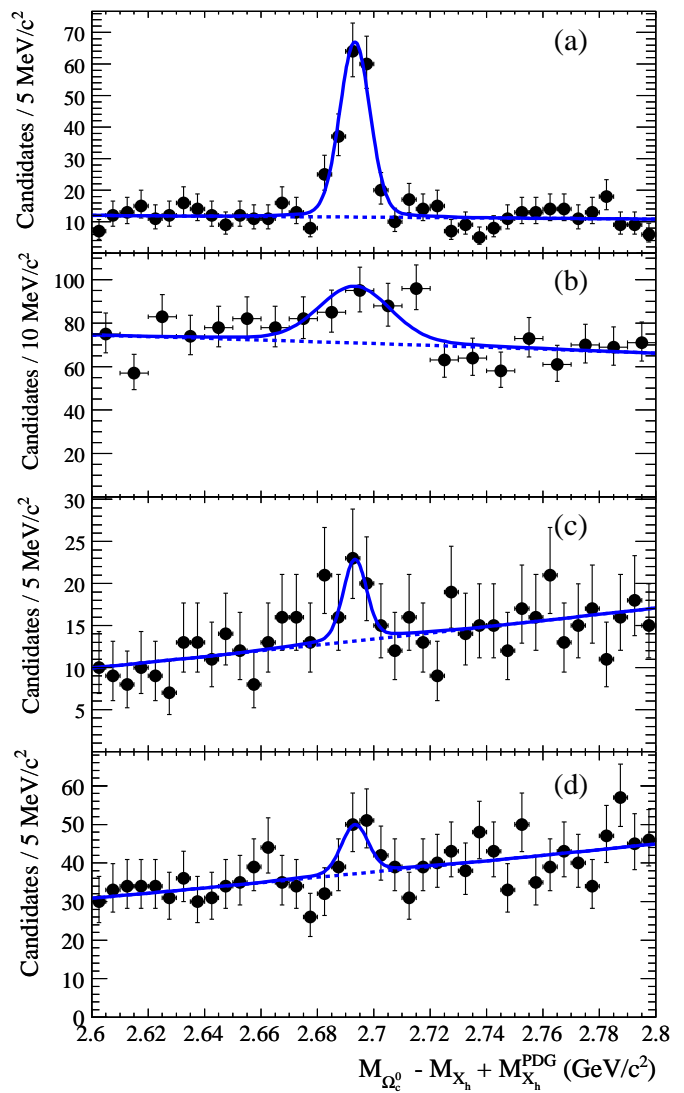}
\caption{ The invariant mass distributions of \ocz\ candidates reconstructed in the \ocz\ decay modes into (a) $\Omega^{-} \pi^{+}$, (b) $\Omega^{-} \pi^{+}\pi^{0}$, (c) $ \Omega^{-} \pi^{+}\pi^{-}\pi^{+}$, and (d) $\Xi^{-} K^{-}\pi^{+}\pi^{+}$. For all of these, we require $x_{p}({\ocz})>$ 0.5. Here $M_{\ocz}$ is the reconstructed mass of the \ocz\ candidates, and X$_{h}$ denotes the daughter hyperon. The points with error bars represent the data, the dashed line represents the combinatorial background and the solid line the sum of signal and background.}
\label{OCZ_PLOT}
\end{center}
\end{figure}

  Fig.~\ref{OCZ_PLOT} shows the reconstructed invariant mass distributions of \ocz\ candidates with $x_{p}({\ocz})>$ 0.5. Clear peaks indicating production of \ocz\ are visible in each of the modes represented in Fig.~\ref{OCZ_PLOT}. The invariant mass resolution is improved by 25$\%$ by using the variable $M_{\Omega^{-}\pi^{+}} - M_{\Omega^{-}} + M_{\Omega^{-}}^{PDG}$, instead of $M_{\Omega^{-}\pi^{+}}$, where $M_{\Omega^{-}}$ is the reconstructed mass of the \omm\ and $M_{\Omega^{-}}^{PDG}$ is the world average mass of the \omm \cite{PDG}. An unbinned extended maximum likelihood (ML) fit is performed to extract the signal yield. For each mode, a double Gaussian function with a common mean is used to fit the signal and a first-order polynomial is used to model the combinatorial background. The mass resolution in each decay mode is obtained from a large sample of MC signal events reconstructed and processed in the same way as data. For the fits shown in Fig.~\ref{OCZ_PLOT}, the widths of the signal lineshapes are fixed to the values from MC simulation. The fit shown in Fig.~\ref{OCZ_PLOT}(a) results in a raw (i.e. uncorrected) yield of $156 \pm 15 ~\stat$ events and a mean mass of $2693.3 \pm 0.6 ~\stat$ \mevcc. For the other three \ocz\ decay modes the mean masses are fixed at 2693.3 \mevcc, and a second-order polynomial is used to model the combinatorial background. The fitted raw yields are $92^{+26}_{-25}~\stat$, $23^{+10}_{-9}~\stat$ and $34^{+15}_{-14}~\stat$ events for O2, O3 and C1 decay modes, respectively.

\begin{figure}[!h]
\includegraphics[height=7.2in]{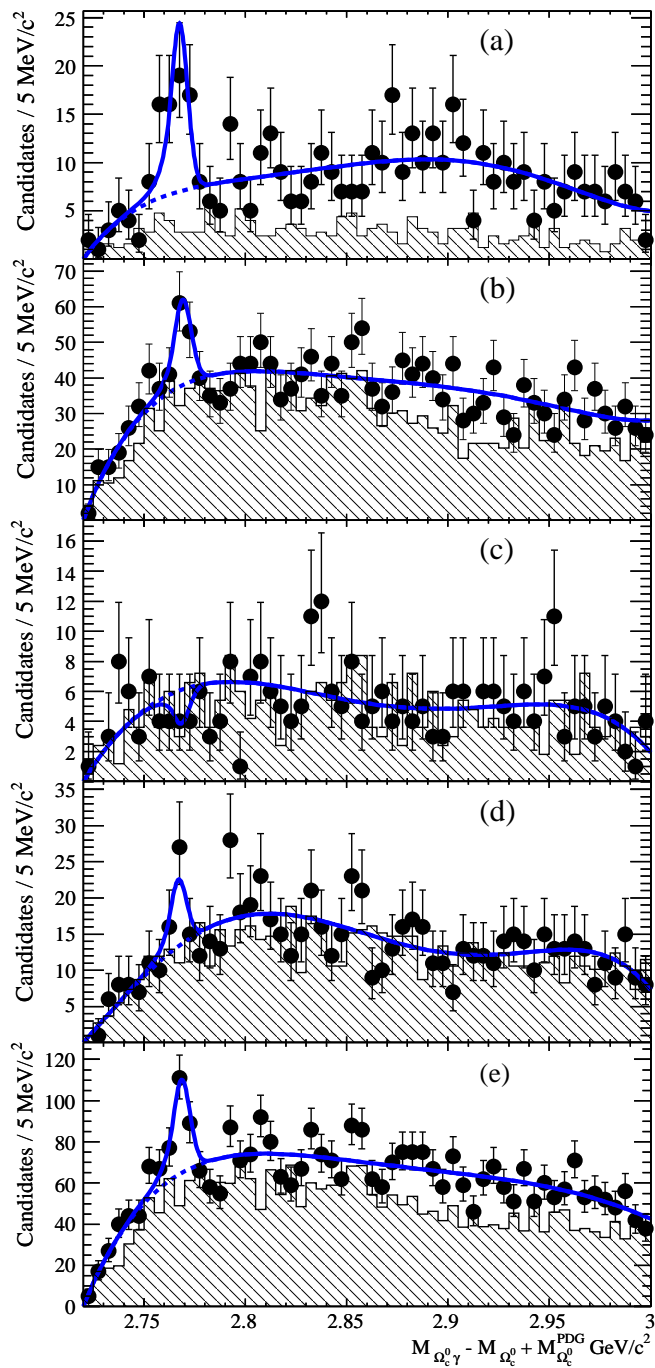}
\caption{ The invariant mass distributions of \ocstg candidates, with \ocz\ reconstructed in the decay modes (a) $ \Omega^{-} \pi^{+}$, (b) $ \Omega^{-} \pi^{+}\pi^{0}$, (c) $\Omega^{-} \pi^{+}\pi^{-}\pi^{+}$, (d) $\Xi^{-} K^{-}\pi^{+}\pi^{+}$, and (e) for the combined decay modes (O1, O2, O3 and C1). For all of these, we require $x_{p}({\ocs}) >$ 0.5. Here $M_{\ocz\g}$ is the reconstructed mass of the \ocs candidates, and $M_{\ocz}$ is the reconstructed mass of the \ocz. The points with error bars represent the data, the dashed line represents the combinatorial background and the solid line the sum of signal and background. The shaded histograms represent the mass distribution expected from the mass sideband of \ocz. }
\label{OCS_PLOT}
\end{figure}

  For \ocs\ candidate selection, we require $x_{p}({\ocs})>$ 0.5 but make no direct cut on $x_{p}({\ocz})$. The invariant mass distributions of \ocstg\ candidates are shown in Fig.~\ref{OCS_PLOT}. The invariant mass resolution is improved by $\approx$ 40\% by using the variable $M_{\ocz\g} - M_{\ocz} + M_{\ocz}^{PDG}$, instead of $M_{\ocz\g}$, where $M_{\ocz}$ is the reconstructed mass of the \ocz\ and $M_{\ocz}^{PDG}$ is the world average mass of the \ocz\ (2697.5 \mevcc)~\cite{PDG}. A clear peak from $\ocs \to\ocz\g$ ($\ocz\to\omm\pip$) production can be seen in Fig.~\ref{OCS_PLOT}(a). The scaled \ocz\ sidebands which are also shown in Fig.~\ref{OCS_PLOT}, show no peak in the mass distribution. The distribution is fitted with the Crystal Ball function~\cite{cball} to model the signal and the product of a fourth-order polynomial and a two-body phase space function~\cite{PDG} to model the combinatorial background. The signal shape parameters are fixed to the values found from MC simulation except for the mean of the distribution. The invariant mass resolution is 4.0 \mevcc. The fit results in $\Delta M$ = $69.9 \pm 1.4 ~\stat$ \mevcc and a raw yield of $39^{+10}_{-9} ~\stat$ events. The fit is superimposed on Fig.~\ref{OCS_PLOT}(a). The signal observed for \ocstg\ ~(\octpi)~ corresponds to a significance of 4.2 standard deviations ($\sigma$) including the systematic uncertainty on the observed yield. The significance is derived from $\sqrt{2{\textnormal{ln}}(L_{max}/L_{0})}$, where $L_{max}$ and $L_{0}$ are the likelihoods for fits with and without a resonance peak component, respectively. The systematic uncertainty is discussed later. We use a similar fit procedure for O2, O3, and C1 decay modes to extract the signal yields. For O3, $M_{\ocz}$ is fixed to the value obtained from the process O1. The fits result in raw yields of $ 55^{+16}_{-15} ~\stat$, $-5\pm5 ~\stat$, and $20\pm9 ~\stat$ events for O2, O3, and C1, respectively. 

 For all decay modes we determine the ratio of inclusive production cross sections,
\begin{center}
 $R = \displaystyle{\frac{\sigma(\epem \to \ocs X, x_{p}(\ocs)>0.5)}{\sigma(\epem \to \ocz X, x_{p}(\ocz)>0.5)}}$,
\end{center}
 where the scaled momentum of the \ocs\ (\ocz) is required to be greater than 0.5 in the numerator (denominator) cross section. We assume that \BR(\ocs\to\ocz\g) = 100$\%$, and include \ocz\ baryons coming from \ocs\ decay as part of the denominator cross section, provided they satisfy the $x_{p}(\ocz)$ requirement. The relative detection efficiencies ($\epsilon_{\ocs}/\epsilon_{\ocz}$) of the \ocs compared to \ocz within these momentum ranges are estimated from MC simulation and are listed in Table~\ref{OCS_FIT}, along with the results for the cross section ratios $R$. 

 We combine O1, O2 , O3, and C1 and perform a single ML fit. The fit results in $\Delta M$ = $70.8 \pm 1.0 ~\stat$ \mevcc, a raw signal yield of $105 \pm 21 ~\stat$ events, with a significance of 5.2$\sigma$ (including systematic uncertainty), and a ratio $~R ~= ~1.01~\pm 0.23 ~\stat$. This procedure weights the individual decay modes by the observed number of \ocz\ baryons in the data, and results in the minimum overall error on the combined value of $R$. The results are summarized in Table~\ref{OCS_FIT}.

\begin{table*}[!htb]
\caption{ The mass difference, $\Delta{M}=M_{\ocs}-M_{\ocz}$ (\mevcc), the fitted signal yield, $Y$ (events), the \ocs signal significance, $S$ (in $\sigma$), the relative detection efficiency, $\epsilon_{\ocs}$/$\epsilon_{\ocz}$, and the ratio of inclusive production cross sections, $R$, as defined in the text. The first uncertainty is statistical, and the second is systematic.}
\begin{center}
\tabcolsep=4mm
\renewcommand{\arraystretch}{1.4}
\begin{tabular}{ l c c c c c c } \hline\hline
Decay mode& $\Delta M$ (\mevcc) & $Y$ (Events) & $S$ ($\sigma$)  & $\epsilon_{\ocs}$/$\epsilon_{\ocz}$ & $R$ \\ \hline
 O1  & $69.9 \pm 1.4 \pm 1.0$ & $39 ^{+10}_{-9} \pm 6$& 4.2  & 0.35 & $0.71^{+0.19}_{-0.18}\pm 0.11$\\ 
 O2  & $71.8 \pm 1.3 \pm 1.1$ & $55^{+16}_{-15}\pm 6$& 3.4& 0.34 & $1.76^{+0.71}_{-0.69}\pm 0.21$\\ 
 O3  &  69.9 (fixed) & $-5\pm 5 \pm 1$ & -& 0.33&  $-0.66^{+0.74}_{-0.66} \pm 0.13$\\ 
 C1  & $69.4^{+1.9}_{-2.0} \pm 1.0$ & $20 \pm 9 \pm 3$ & 2.0 & 0.35 & $1.70^{+1.02}_{-1.00}\pm 0.34$\\ 
Combined  & $70.8 \pm 1.0 \pm1.1$ & $105 \pm 21 \pm6$ & 5.2 & 0.34 & $1.01 \pm 0.23 \pm 0.11$\\ \hline\hline
\end{tabular}
\end{center}
\label{OCS_FIT}
\end{table*}
\renewcommand{\arraystretch}{1.0}

Several sources of systematic uncertainty in the fitted signal yields
are considered. The largest uncertainties arise from the
fits to the mass spectra. These are estimated by repeating the fits,
varying the fixed parameters of the fitted signal functions by $\pm1$
standard deviation and varying the functional parametrization of the
background. The systematic uncertainty on the yield from the combined
\ocs\ modes is $6\%$. The systematic uncertainty on
$\Delta M$ is dominated by the photon energy scale and is 1.5$\%$. This is estimated from the distribution of reconstructed masses of low-energy neutral pions. The uncertainty in the fitting procedure leads to a systematic uncertainty of
$11\%$ on the ratio $R$, measured from the combined modes.
There are also systematic uncertainties of 1.8$\%$
from the photon reconstruction efficiency, and 1.4$\%$ due to the limited MC sample size. The uncertainties
from tracking, particle identification, selection of intermediate hyperon
candidates, daughter branching fractions~\cite{PDG} and luminosity approximately cancel in the ratio, since the \ocs\ analysis uses the same selection and
data sample as the \ocz\ analysis. The sensitivity to fragmentation modeling is negligible. A possible additional uncertainty arises from multiple candidates found in $\approx 10\%$ of the events in the data, usually due to a common hyperon combined with alternative particles from the rest of the event to form \ocs\ candidates. These are uniformly distributed in $M_{\ocs}$ and are hence absorbed into the background parametrization, with no evidence for multiple candidates peaking in mass.

 In summary, we report the first observation of an excited singly-charmed baryon \ocs\ ($css$) decaying to \ocz\ and a photon, with a significance of 5.2$\sigma$, and measure the mass difference between \ocs\ and \ocz\ to be $\Delta M~=~70.8~\pm~1.0 ~\stat \pm 1.1 ~\syst$ \mevcc. This is consistent with the theoretical prediction in [2,~4-11] and below that described in~\cite{1}. We also measure the ratio of inclusive production cross sections,
\begin{center}
$R  = 1.01 \pm 0.23~\stat \pm 0.11~\syst$.
 \end{center}

We are grateful for the excellent luminosity and machine conditions
provided by our \pep2\ colleagues, 
and for the substantial dedicated effort from
the computing organizations that support \babar.
The collaborating institutions wish to thank 
SLAC for its support and kind hospitality. 
This work is supported by
DOE
and NSF (USA),
NSERC (Canada),
IHEP (China),
CEA and
CNRS-IN2P3
(France),
BMBF and DFG
(Germany),
INFN (Italy),
FOM (The Netherlands),
NFR (Norway),
MIST (Russia), and
PPARC (United Kingdom). 
Individuals have received support from the
Marie Curie EIF (European Union) and
the A.~P.~Sloan Foundation.

\end{document}